\documentclass[aps,prd,amsmath,amssymb]{revtex4}
\usepackage{amsmath, amsthm, amssymb}
\usepackage{graphicx}% Include figure files
\usepackage{dcolumn}% Align table columns on decimal point
\usepackage{bm}% bold math

\begin{document}

%\preprint{APS/123-QED}

%\documentclass[12pt]{article}

%\usepackage{amssymb}
%\usepackage{graphicx}
%\usepackage{enumerate}

%\newcommand{\sect}[1]{\section{#1}\setcounter{equation}{0}}
%\def\theequation{\thesection.\arabic{equation}}
%\setlength{\topmargin}{0pt}
%\setlength{\oddsidemargin}{10pt}
%\setlength{\evensidemargin}{10pt}
%\setlength{\textwidth}{6.2 truein}
%\setlength{\textheight}{8 truein}
%\def\thesection{\arabic{section}}

%\begin{document}

%\twocolumn

\def\beq{\begin{equation}}
\def\eeq{\end{equation}}
\def\bea{\begin{eqnarray}}
\def\eea{\end{eqnarray}}
\def\ben{\begin{enumerate}}
\def\een{\end{enumerate}}
\def\la{\langle}
\def\ra{\rangle}
\def\a{\alpha}
\def\b{\beta}
\def\g{\gamma}\def\G{\Gamma}
\def\d{\delta}
\def\e{\epsilon}
\def\phi{\varphi}
\def\k{\kappa}
\def\l{\lambda}
\def\m{\mu}
\def\n{\nu}
\def\o{\omega}
\def\p{\pi}
\def\r{\rho}
\def\s{\sigma}
\def\t{\tau}
\def\L{{\cal L}}
\def\S{\Sigma }
\def\gsim{\; \raisebox{-.8ex}{$\stackrel{\textstyle >}{\sim}$}\;}
\def\lsim{\; \raisebox{-.8ex}{$\stackrel{\textstyle <}{\sim}$}\;}
\def\gtrsim{\gsim}
\def\lessim{\lsim}
\def\loc{{\rm local}}
\def\vm{v_{\rm max}}
\def\bh{\bar{h}}
\def\del{\partial}
\def\nab{\nabla}
\def\half{{\textstyle{\frac{1}{2}}}}
\def\fourth{{\textstyle{\frac{1}{4}}}}

\title{Two-dimensional gravity with a dynamical aether}

\author{Christopher Eling}
 \email{cteling@physics.umd.edu}
 \author{Ted Jacobson}
\email{jacobson@umd.edu}
 \affiliation{Department of Physics, University of Maryland\\ College Park, MD 20742-4111 USA}

\begin{abstract}

We investigate the two-dimensional behavior of gravity coupled to a
dynamical unit timelike vector field, i.e. ``Einstein-aether
theory". The classical solutions of this theory in two dimensions
depend on one coupling constant. When this coupling is positive the
only solutions are (i) flat spacetime with constant aether, (ii) de
Sitter or anti-de Sitter spacetimes with a uniformly accelerated
unit vector invariant under a two-dimensional subgroup of $SO(2,1)$
generated by a boost and a null rotation, and (iii) a non-constant
curvature spacetime that has no Killing symmetries and contains
singularities. In this case the sign of the curvature is determined
by whether the coupling is less or greater than one. When instead
the coupling is negative only solutions (i) and (iii) are present.
This classical study of the behavior of Einstein-aether theory in
1+1 dimensions may provide a starting point for further
investigations into semiclassical and fully quantum toy models of
quantum gravity with a dynamical preferred frame.

\end{abstract}

%\pacs{}
\maketitle
%\tableofcontents

Theories of gravity in two spacetime dimensions have provided useful
toy models for the investigation of issues such as black hole
evaporation and information loss, singularities, and quantization of
gravity, in a setting where some of the technical problems that
arise in four dimensions are absent. In two dimensions the
Einstein-Hilbert action is topological, so something else must be
used. An early proposal was the Jackiw-Teitelboim model \cite{JT},
for which the equations of motion imply the metric has a fixed
constant curvature specified by a parameter in the Lagrangian. A
generalization of this idea is dilaton gravity
\cite{Grumiller:2002nm}, which is a class of models that
incorporates a scalar field in addition to the metric. These
theories possess no local degrees of freedom, but there exist for
example black hole solutions \cite{Grumiller:2002nm,
Brown:1986nm,Frolov:1992xx,Callan:1992rs,Achucarro:1993fd}, and when
coupled to matter fields the theories acquire local dynamics.

In this paper we examine the two dimensional version of
 ``Einstein-aether"  theory,
which is a generally covariant theory consisting of general
relativity coupled to a dynamical unit timelike vector field $u^a$.
Our motivation is that, like other two dimensional models, this
might provide a useful setting in which to study aspects of quantum
gravity. A special feature of this theory is the presence of the
unit timelike vector field, which might ameliorate or modify the
problem of time in the canonically quantized setting. The vector
also defines an intrinsic preferred frame with which for example the
impact of Lorentz violation on black hole evaporation could be
studied. The two dimensional case provides the simplest context in
which to begin examining this idea. It may also be relevant to the
spherical reduction of the higher dimensional theory. For a review
of motivations, history and status of the 3+1 dimensional
Einstein-aether theory as of 2004 see \cite{Eling:2004dk} and the
references therein.  A more recent brief review with further
references is given in the introduction of \cite{Eling:2006df}.

A unit vector field in two dimensions has only one degree of
freedom, so in this respect is similar to a dilaton field. Like the
dilaton, the presence of the vector field renders the theory
nontrivial, but still with no local degrees of freedom. However,
Einstein-aether theory seems to provide a different two dimensional
gravity model than any previously considered. It possesses both
constant and non-constant curvature solutions.  Unlike the
Jackiw-Teitelboim model the constant curvature is not specified a
priori by the action. In this regard it is similar to
two-dimensional unimodular gravity \cite{unimod}, but unlike in
unimodular gravity the {\it sign} of the curvature scalar is
determined by the action. Also it has the unit vector field, which
defines in each solution a locally preferred frame. The gradient of
a dilaton field also defines a vector, and so we looked for a
correspondence between unimodular dilaton gravity and ae-theory in
two dimensions, but so far have not identified any precise mapping
between the theories.

\section{1+1 dimensional action}
\label{reducedaction}

The action for Einstein-aether theory in $n$ dimensions is
\beq S[g_{ab},u^a,\lambda] = \frac{-1}{16\pi G}\int d^nx\,
\sqrt{-g}\Bigl(R+L_u+ \lambda(g_{ab}u^a u^b - 1)\Bigr), \label{action}
\eeq
where $R$ is the Ricci scalar, $\lambda$ is a Lagrange
multiplier enforcing the unit timelike
constraint on $u^a$, and the aether Lagrangian is defined by
\beq L_u = c_1(\nabla_a u_b)(\nabla^a u^b) +c_2(\nabla_a u^a)^2
+c_3(\nabla_a u_b)(\nabla^b u^a) +c_4(u^a \nabla_a u^b)(u^c \nabla_c
u_b) \label{lagrangian} \eeq
where the $c_i$ are dimensionless coupling constants, and
$\lambda$ is a Lagrange multiplier. Here we use the signature $({+} {-} \cdots {-})$.
This action includes all generally covariant terms with up two
two derivatives (not including total divergences) that can be constructed from
a metric and a unit vector field.

In two-dimensional spacetime the variation of the Einstein-Hilbert
term $\sqrt{-g}R$ is a total divergence. The aether part of the
action is non-trivial, but only two of the terms are independent. To
see this we express the covariant derivative $\nabla_a u_b$ in the
orthonormal basis $\{u^a, s^a\}$, where $s^a$ is a unit spacelike
vector orthogonal to $u^a$.  It follows from $u^a u_a = -s^as_a=1$
and $u^a s_a = 0$ that $0=u^b \nabla_a u_b =s^b \nabla_a s_b =u^b
\nabla_a s_b+s^b \nabla_a u_b$ everywhere. Using these relations, we
find that when the unit constraint is satisfied the covariant
derivatives take the form
\bea \nabla_a u_b &=& A s_a s_b + B u_a s_b, \label{uexp}\\
 \nabla_a s_b &=&  A s_a u_b+B u_a u_b. \label{sexp}\eea
where $A$ and $B$ are generically spacetime functions. Using
(\ref{uexp}) we obtain
\bea (\nabla_a u_b)(\nabla^a u^b) &=& A^2-B^2\label{dusq} \\
(\nabla_a u_b)(\nabla^b u^a) &=& A^2 \\
(\nabla_a u^a)^2 &=& A^2\label{div}\\
(u^a \nabla_a u^b)(u^c \nabla_c u_b) &=& -B^2  \\
F_{ab}F^{ab} &=& -2B^2, \label{Fsq}\eea
where $F_{ab} =  \nabla_{a} u_{b}-\nabla_b u_a$. These expressions
may be substituted into the Lagrangian (\ref{lagrangian}) without
changing the equations of motion, since the Lagrange multiplier term
in the action (\ref{action}) implies that the equations of motion
are equivalent to the condition that the action be stationary only
with respect to variations of $u^a$ that preserve the constraint
$u^au_a=1$. On the constraint surface the Lagrangian is thus given
by
\beq L_u=c_{123}A^2 - c_{14}B^2, \label{LAB}\eeq
where $c_{123}\equiv c_1+c_2+c_3$ and $c_{14}\equiv c_1+c_4$.
Using (\ref{div}) and (\ref{Fsq}) the Lagrangian can
therefore be written for example as
 \beq L_u = \half c_{14} F^{ab} F_{ab} + c_{123} (\nabla_a
u^a)^2, \label{reduced}\eeq
without any loss of generality in the theory.

If either $c_{14}$ or $c_{123}$ vanishes the theory is
under-deterministic. In particular, if $c_{123}=0$ then obviously
any metric and unit vector satisfying $F_{ab}=0$ obey all the field
equations with $\l=0$. If instead $c_{14}=0$ then then any metric
and unit vector satisfying $\nabla_a u^a=0$ obey all the field
equations with $\l=0$. (In fact, these include all solutions.) We
thus assume in the remainder of this paper that neither $c_{14}$ nor
$c_{123}$ is zero. The classical equations of motion then depend
only on the one combination $c_{123}/c_{14}$ of the coupling
coefficients.

The action can be further simplified by a field redefinition of the
form
\bea g_{ab} &=& g'_{ab}+(\s-1)u'_a u'_b \nonumber\\
u^a &=& \s^{-1/2} u'^a. \label{redef}\eea
Here the coefficient $\s$ must be positive in order to preserve
Lorentzian signature. This redefinition preserves the general form
of the action given in (\ref{action}), the overall effect being only
a change of the coupling constants,
\beq S[g_{ab},u^a,c_i]=S[g'_{ab},u'^a,c'_i(c_i,\s)].\eeq
The relation between $c'_i$ and $c_i$ was found by
Foster\cite{Foster:2005ec}, and is most usefully specified by
certain linear combinations that have simple transformation
behavior,
\bea
c'_{14} &=& c_{14}\\
c'_{123} &=& \s^{-1} c_{123}\\
c'_{13}-1 &=& \s^{-1} (c_{13}-1)\label{c13}\\
c'_1-c'_3-1 &=& \s(c_1-c_3-1). \label{cc'}\eea
This result applies in any spacetime dimension. The $c'_i$ receive
contributions from the $R$ term in the action which appear in
(\ref{c13}) and (\ref{cc'}) via the $c_i$--independent terms.
However in 1+1 dimensions these contributions cannot affect the
equations of motion since as noted above the variation of the $R$
term is a total divergence. Indeed, the Lagrangian (\ref{reduced})
depends only on the two combinations $c_{14}$ and $c_{123}$ whose
transformation has no $c_i$--independent terms.

Since $c_{14}$ is invariant and $c_{123}$ simply scales by the
nonzero factor $\s^{-1}$, no field redefinition will make one of the
terms in (\ref{LAB}) or (\ref{reduced}) vanish. On the other hand,
the choice $\s = c_{123}/c_{14}$ (which is allowed as long as it
is positive) will produce $c'_{14}=c'_{123}$, in which case using
(\ref{dusq}) the Lagrangian may be reduced to just one term of the
original four-term Lagrangian (\ref{lagrangian}),
\beq L_{u,{\rm reduced}} = c'_{14} (\nabla_a u_b)(\nabla^a u^b)
=c'_{14}\bigl(F^{ab}F_{ab}+ (\nabla_a u^a)^2). \label{reduced2}\eeq
In terms of the new fields the classical equations of motion are
thus totally independent of the coupling parameters. We shall obtain
the solutions for the general case when $c_{123}/c_{14}$ is positive
by applying the field redefinition to solutions of this reduced
theory.

\section{Field equations and solutions}
\label{fieldequations}

In this section we will study the general two-dimensional theory
described by the action with Lagrangian (\ref{reduced}), to obtain
the field equations and the curvature of their solutions.
%For this
%general case we use the primed couplings $c'_i$, reserving the
%unprimed couplings for the reduced case (\ref{reduced2}) obtained by
%field redefinition.
The Lagrangian takes the form
 \beq L_u = c_{14}\left( \half F^{ab} F_{ab} + \b (\nabla_a
u^a)^2\right), \label{reduced'}\eeq
where
\beq \b=c_{123}/c_{14}. \label{beta} \eeq
The equations of motion depend on the couplings only through
the combination $\b$ defined in (\ref{beta}).
Varying with respect to the inverse metric
$g^{ab}$ and the covariant aether vector $u_a$ as independent
field variables, we find the metric field equation
\bea  F_{am} F_b{}^{m} &-&\half g_{ab}(\half  F^{ab} F_{ab} - \b
(\nabla_c u^c)^2 - 2 \b u^m \nabla_m(\nabla_c u^c))\nonumber\\&-&2\b
u_{(a}\nabla_{b)}\nabla_c u^c+ \l u_a u_b =0 \label{metriceqn} \eea
and the aether field equation
\beq  \nabla_b F^{ba} + \b \nabla^a(\nabla_c u^c) - \l u^a = 0,
\label{aeeqn} \eeq
where $\l$ is a re-scaled Lagrange multiplier ({\it cf.} eqn.
(\ref{action})). These amount to three equations from
(\ref{metriceqn}) and two from (\ref{aeeqn}). The $u^a$ component of
the latter determines $\l$. Using the expansion of the covariant
derivatives (\ref{uexp},\ref{sexp}) to project out the various
components of the field equations we find that the remaining four
equations are equivalent to
\bea &uA=f \qquad &sA=0,\nonumber \\
&uB=0\qquad &sB=\b f  \label{feqns}\eea
where (for example) $uA\equiv u^m \nabla_m A$, and
\beq f \equiv \half(A^2 - \b^{-1}B^2). \label{f}\eeq

The equations (\ref{feqns}) are extremely restrictive, and there are
just two types of solutions. In the first type both $A$ and $B$ are
constant and related by $B^2=\b A^2$ so that $f=0$. In the second
type of solution the gradients $\nabla_a A$ and $\nabla_b B$ are
both non-zero and independent. In this case $A$ and $B$ may be used
as coordinates, so one can immediately write the unique solution,
\beq u=f\partial_A\qquad\mbox{and}\qquad s=\b f\partial_B.\label{ABusolution} \eeq
The inverse metric is given by $g^{ab}=u^au^b - s^a s^b$, hence for
this solution the line element is
\beq ds^2=\frac{4}{(\b A^2-B^2)^2}(\b^2 dA^2 - dB^2).
\label{ABsolution}\eeq

The scalar curvature $R$ completely characterizes the curvature in
two-dimensions. Using the relation
\beq R= 2u^a(\nabla_b \nabla_a-\nabla_a \nabla_b)u^b\eeq
which is valid in two-dimensions, and making use of
(\ref{uexp}) and (\ref{sexp}), we find
\beq R = 2(B^2-A^2+u A + s B). \label{R} \eeq
When the field equations (\ref{feqns}) are satisfied the scalar
curvature is thus given by
\beq R = (\b - 1) A^2 + (1-\b^{-1}) B^2. \label{Rsol}\eeq
The solutions with constant $A$ and $B$ thus have constant
curvature. Being two dimensional, they are therefore either
Minkowski, de Sitter, or anti-de Sitter space, and have three
independent Killing vectors. The solution (\ref{ABsolution}) has
non-constant curvature unless $\b=1$, in which case it is flat. For
$\b\ne1$ it can be shown that (\ref{ABsolution}) has no Killing
vectors. For $\b>1$ the curvature scalar is positive, for $0<\b<1$
it is negative, and for $\b<0$ it is indefinite.

In the case $\b<0$ the function $f$ defined in (\ref{f}) can only
vanish when both $A$ and $B$ vanish, hence the only solution with
constant $A$ and $B$ is the one with $A=B=0$. In this solution the
metric is flat, and according to (\ref{uexp}) the vector field $u^a$
is then constant. The only other solution in this case is
(\ref{ABusolution},\ref{ABsolution}). The curvature scalar for this
metric with $\b<0$ is zero on the lines $|B|=|A|(1-\b)/(1-\b^{-1})$,
negative for smaller $|B|/|A|$ and positive for larger $|B|/|A|$. It
vanishes at $A=B=0$, which lies at infinite distance diverging as
$1/A$ on any non-null line $A/B={\rm const}$. There is a curvature
singularity as either $A$ or $B$ goes to infinity, except on the
lines where the curvature vanishes, and the geodesic distance to
this singularity is finite.

In the next section we determine the nature of the solutions
for the special case
$\b=1$, and the following section addresses  the  case $0<\b\ne1$.
Since the Lagrangian with $\b>0$ can be reached by a field
redefinition from the $\b=1$ case, the solutions for general positive $\b$
can be obtained by field redefinition from the $\b=1$ solutions.

\section{$\b=1$: Flat spacetime solutions}
\label{flatsol}

To find the solutions in the case $\b=1$, for which the curvature
vanishes, we adopt null coordinates $(w,v)$, so
\bea ds^2 &=& dw\, dv \label{dwdv}\\
u &=& F \partial_w +  F^{-1} \partial_v \eea
where $F(w,v)$ is to begin with an arbitrary function. The
functions $A$ and $B$ defined in (\ref{uexp}) and (\ref{sexp}) are then
given by
\bea A &=& -F_{,w} + F^{-2} F_{,v}\nonumber\\
B &=& F_{,w} + F^{-2} F_{,v}. \label{AB}\eea
Using the field equations in the form (\ref{feqns}) with $\b=1$,
we find that $uB+sA=0$ implies $(\log F)_{,wv} = 0$,
$(u+s)(A-B)=0$ implies $F_{,ww}=0$,  and $(u-s)(A+B)=0$ implies
$(1/F)_{,vv}=0$. It follows that the general solution is
\beq F = (a + bw)/(c+dv), \eeq
where $a,b,c,d$ are constants. Thus there are four classes of
solutions for $u^a$, corresponding to whether or not the constants
$b$ and $d$ vanish. Up to constant coordinate shifts and opposite
scalings of $w$ and $v$ (which preserve the Minkowski metric
(\ref{dwdv})) these four solutions have $(w,v)$ components
\bea &(1,1) \label{usol1}\\
&(k w,(kw)^{-1})  \label{usol2}\\
&((kv)^{-1},kv) \label{usol3}\\
&(w/v, (w/v)^{-1})\label{usol4}\eea
where $k$ is a constant with dimensions of
inverse length that sets a physical scale for the solution.

\begin{figure}
 \includegraphics[angle=0,width=6cm]{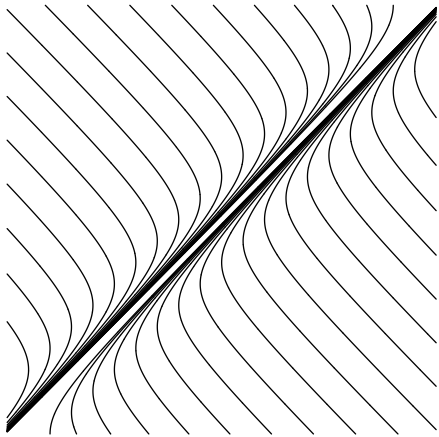}\\
\caption{\label{uflat1} Plot of the flow lines of (\ref{usol2}) in
Minkowksi space with Cartesian coordinates $t$ (increasing
vertically) and $x$ (increasing toward the right). The $u^a$ field
approaches the null vector $\partial_v$ along the line $w=t-x=0$,
hence must be infinitely stretched there in order to maintain the
unit constraint.}
\end{figure}

The first solution  (\ref{usol1}) is simply a constant vector field
covering the entire Minkowski spacetime. In this solution $A$ and
$B$ both vanish. The second and third solutions (\ref{usol2}) and
(\ref{usol3}) are equivalent to each other with the roles of $w$ and
$v$ reversed. For the solution (\ref{usol2}) we use (\ref{AB}) to
find that $A=-B=-k$, so these are solutions of the type with $A$ and
$B$ constant. In the solution (\ref{usol2}) the vector field $u^a$
is non-singular in regions covering one-half of the flat Minkowski
manifold, either $w>0$ or $w<0$. Along the line $w=0$ the vector
$u^a$ becomes infinitely stretched in order to maintain unit norm as
it aligns with the null vector $\partial_v$. There is a similar
divergence as $w\rightarrow\pm\infty$ where $u^a$ aligns with
$\partial_w$. The flow lines of $u^a$ are the level curves of a
function $\Phi$ with $u\Phi =kw \Phi _{,w}+(kw)^{-1}\Phi _{,v}=0$,
which is satisfied by $\Phi =v + (k^2w)^{-1}$. Thus the flow lines
are given by
 \beq v+(k^2w)^{-1}={\rm const.} \eeq
These curves are hyperbolae, as can also be seen from the fact
that the acceleration vector $u^a\partial_a u^b$ is given in
$(w,v)$ components by $kw
\partial_w   (kw, (kw)^{-1})=(k^2w, -w^{-1})$, which has the
constant squared norm $-k^2$. A plot showing these flow lines in a
part of the Minkowski space is shown in Fig. \ref{uflat1}.

The solution (\ref{usol2}) is further characterized by its
symmetries. Being flat, the metric has two translational
symmetries generated by the Killing vectors $\partial_w$ and
$\partial_v$, and one boost symmetry generated by
$w\partial_w-v\partial_v$. The vector field $u^a$ is clearly
invariant under $\partial_v$, since its components depend only
upon $w$. It is also invariant under the boost Killing vector:
$[w\partial_w-v\partial_v,kw\partial_w +(kw)^{-1}\partial_v]=0$.
The commutator of these two Killing vectors that commute with
$u^a$ is
\beq [w \partial_w- v\partial_v, \partial_v] = \partial_v.
\label{algebra1}\eeq
They generate a non-abelian sub-algebra of the Poincare algebra in
1+1 dimensions. This sub-algebra is isomorphic to the algebra of
the affine group $A(1)$ of translations and scalings in one
dimension. It will re-appear in the next section as a sub-algebra
of the 2+1 dimensional Lorentz group when we relate this solution
to a constant curvature one via a field redefinition.

\begin{figure}
 \includegraphics[angle=0,width=6cm]{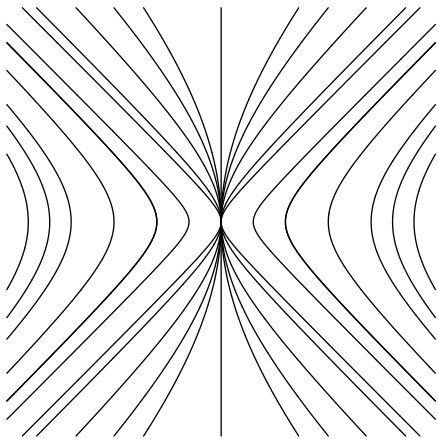}\\
\caption{\label{uflat2} Plot of the flow lines of (\ref{usol4}) in
Minkowksi space with Cartesian coordinates $t$ and $x$. Here the
aether is singular along the lines $w=t-x=0$ and $v=t+x=0$.}
\end{figure}

For the fourth solution (\ref{usol4}) we again use (\ref{AB}) to
find that now
\bea
A&=&-w^{-1}-v^{-1}\nonumber\\
B&=&-w^{-1}+v^{-1}.\label{AB4} \eea These are not constant, so this
solution corresponds to the solution
(\ref{ABusolution},\ref{ABsolution}) with $\b=1$. Though not
obviously flat, this line element is evidently related to the
Minowski metric by the coordinate transformation (\ref{AB4}). The
vector field $u^a$ in this solution is singular on both lines $w=0$
and $v=0$, and stretches infinitely as either $v$ or $w$ goes to
infinity. The solution is thus regular in any of the four wedges
$(w>0,v>0)$, $(w<0,v<0)$, $(w>0,v<0)$, and $(w<0,v>0)$. The first
two are related by time reflection and the last two by space
reflection, but the first pair is physically distinct from the
second pair. The flow lines in this case are the level curves of a
function $\Phi $ with $u\Phi =(w/v)\Phi _{,w} + (v/w)\Phi _{,v}=0$,
which is satisfied by $\Phi =w^{-1}-v^{-1}$. The flow lines are
therefore given by
\beq w^{-1}-v^{-1}={\rm const.} \eeq
These curves do not have constant acceleration. A plot showing
them in a part of the Minkowski space is shown in Fig.
\ref{uflat2}. Unlike the previous case, this $u^a$ field commutes
with none of the Killing vectors of the flat metric.

This completes our analysis of the solutions in the special case
when the coupling constants satisfy $\b=1$, for which the metric is
flat. Next we turn to determining the solutions for general $\b>0$.

\section{$0<\b\ne1$ solutions} \label{generalsol}

To obtain the general solutions for the theory when $0<\b\ne1$ we
use the field redefinition (\ref{redef}) with $\s=\b$,
\bea g_{ab} &=& \eta_{ab}+(\b-1)u'_a u'_b \nonumber\\
u^a &=& \b^{-1/2}~ u'^a. \label{redef1}\eea
If $(\eta_{ab},u'^a)$ is a solution to the theory with $\b=1$ then
$(g_{ab},u^a)$ is a solution with arbitrary positive
$\beta=c_{123}/c_{14}$. Conversely, every solution of the
$\b\ne1$ theory can be obtained in this way. We apply this method
to the three different types of solutions found in the previous
section.

Under the field redefinition (\ref{redef1}) the new line
element is
\beq ds^2 = \fourth(\b-1)(u'^w)^{-2} dw^2 + \half(\b+1) dw dv +
\fourth(\b-1)(u'^w)^2 dv^2 \label{primedmetric}\eeq
and the new aether is
\beq u= \b^{-1/2} (u'^w \partial_{w} + u'^v \partial_{v}),
\label{primedu}\eeq
where we use the same coordinates $(w,v)$ to describe the new
solution as we used for the flat one. The determinants of both the
primed (flat) and unprimed metrics are constant in these
coordinates, hence we have $\nabla_a u^{a}=\nabla'_a u^{a}$ and
similarly for $s^a$, so using (\ref{uexp}), (\ref{sexp}) and
(\ref{redef1}) we find
\beq A=\b^{-1/2}A'\qquad\mbox{and}\qquad B=B'.\label{AA'}\eeq
The curvature
(\ref{Rsol}) of the new metric is therefore given by
\beq R = (1-\b^{-1})(A'^2+B'^2), \label{R'sol}\eeq
where $A'$ and $B'$ are those of the primed, flat solution (which
had $\b=1$).

For the constant vector field solution (\ref{usol1}) the primed
metric components remain constant, as do those of the aether, so
after the field redefinition we still have the trivial solution of
a constant aether in a flat spacetime after the field redefinition.

In the next two subsections we consider the solutions obtained by
field redefinition from the other two types of solutions, first
(\ref{usol4}) and next (\ref{usol2}) and (\ref{usol3}).

\subsection{Non-constant curvature solution}
\label{non-constant}

Using (\ref{primedmetric}) with the primed solution (\ref{usol4}) we find
\beq ds^2 = \fourth(\b-1)(w/v)^{2} dw^2 + \half(\b+1) dw dv +
\fourth(\b-1)(v/w)^2 dv^2 \label{nonconstcurvmetric}\eeq
As we saw in (\ref{AB4}) $A'$ and $B'$ are not constant for this
$u'$, hence according to (\ref{AA'}) neither are $A$ and $B$, so
this solution corresponds again to the non-constant curvature
solution (\ref{ABsolution}). The scalar curvature (\ref{R'sol}) of
the new metric (\ref{nonconstcurvmetric}) is given, according to
(\ref{R'sol}) and (\ref{AB4}), by
\beq R = 2 (1-\b^{-1}) (w^{-2}+v^{-2}). \label{R'4}\eeq
As discussed in Section \ref{flatsol},
none of the flat-spacetime Killing vectors commute
with this $u'^a$, from which it follows that
no Killing vector of $g_{ab}$ could commute with
$u^a$. Moreover, in fact $g_{ab}$ has no Killing vectors at all,
as mentioned previously.

When $w$ or $v$ vanishes the metric (\ref{nonconstcurvmetric}) has a
curvature singularity. In the same limits $u^a$ aligns with either
$\partial_{w}$ or $\partial_{v}$, which are null vectors when
respectively $w$ or $v$ equals zero. Thus, for this solution, the
scalar curvature becomes singular exactly on the horizons where
$u^a$ must be infinitely stretched. As in the flat case discussed
above there are two distinct regular solutions (up to time or space
reflection), corresponding to the coordinate ranges $w,v>0$ or $w<0,
v>0$. Approaching the singularity at $w=0$ along a line of constant
$w+v>0$, the distance diverges logarithmically as $\log w$. If
instead we fix $w/v$ and go out to infinite values of $w$ and $v$
the curvature approaches zero, and the distance diverges linearly in
$w$. On the other hand if we fix $w$ and go out to infinite $v$ the
curvature approaches a constant  proportional to $w^{-1}$ and the
distance diverges as $\log v$.

\subsection{Constant curvature solutions}
\label{constantcurv}

\begin{figure}
 \includegraphics[width=10cm,height=5cm,angle=270]{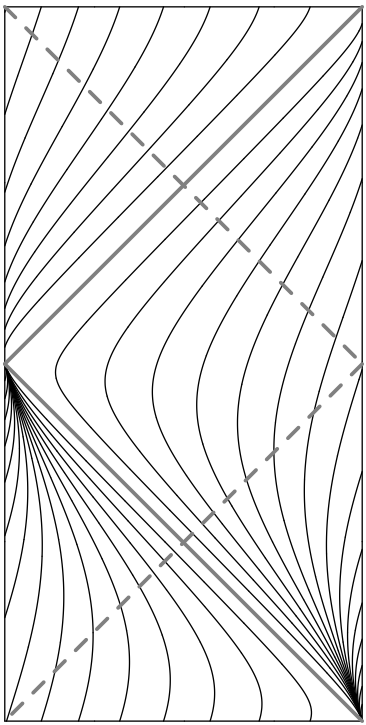}\\
\caption{\label{uAdS} Conformal diagram of 1+1 AdS spacetime with
the flow lines of the aether field. The two vertical lines at left
and right are at null infinity, and the diagram should be continued
infinitely in the vertical direction. The aether is regular in the
Poincar\'e patch bounded by the solid grey lines and null infinity.
The dashed grey lines are the rest of the boost Killing horizon. The
diagram is drawn for the case $\b=10$.}
\end{figure}

Under the field redefinition (\ref{redef1}) the second type of
solution (\ref{usol2}) produces the metric
\beq ds^2 = \fourth(\b-1)(kw)^{-2} dw^2 + \half(\b+1) dw dv +
\fourth(\b-1)(kw)^2 dv^2 \label{constcurvmetric}\eeq
and re-scaled $u^a$
\beq u= \b^{-1/2} (kw \partial_{w} + (kw)^{-1} \partial_{v}).
\label{constcurvu}\eeq
As mentioned in the previous section, (\ref{AB}) implies for the
primed solution $A'=-B'=-k$, so according to (\ref{AA'}) this
solution corresponds to the general type with constants
$A=-\b^{-1/2}k$ and $B=k$, and the scalar curvature (\ref{R'sol}) of
the primed metric (\ref{constcurvmetric}) is
\beq R' = 2 (1-\b^{-1}) k^2, \label{R'}\eeq
The curvature is constant, so the geometry is locally that of
de-Sitter (dS) for $0<\b<1$ and anti-de-Sitter (AdS) space for
$\b>1$ (Recall that we use the metric signature $(+{-})$, so the
scalar curvature for dS is negative while for AdS it is positive.)
The nature of these maximally symmetric spaces is well-known, so to
fully describe these solutions we need only specify the behavior of
the $u^a$ vector field on the dS/AdS background. This behavior is
illustrated for the case of de Sitter and anti-de Sitter spaces in
Fig.~\ref{udS} and Fig.~\ref{uAdS}. In the remainder of this paper
we explore the properties of this solution.

First note that since
\beq u^a \nabla_a u_b = u^a (As_as_b+Bu_as_b) =ks_b\eeq
the magnitude of the acceleration of the flow of $u^a$ with respect
to $g_{ab}$ is constant and equal to $k$, as is that of $u'^a$ with
respect to $\eta_{ab}$. The coordinates $w$ and $v$ in
(\ref{constcurvmetric}) are not null with respect to $g_{ab}$ since
the effect of the $u'_a u'_b$ contribution to (\ref{redef1}) is to
narrow the light cones of the flat metric when $0<\b<1$ and widen
them when $\b>1$. However, $\partial_{v}$ is a null vector when
$w=0$ and similarly $\partial_{w}$ is null when
$w\rightarrow\pm\infty$. From (\ref{constcurvu}), it is clear that
$u^a$ is singular on one of the dS/AdS horizons labelled
by $w=0$, where it is infinitely stretched in order to remain unit
timelike as it approaches a null vector. It is also infinitely
stretched as $w$ approaches $\pm\infty$. The aether is thus regular
in either of the two coordinate patches $w>0$ or $w<0$. It is not
immediately clear to which regions of dS/AdS these patches
correspond. We shall address this shortly with the help of new
coordinates better adapted to the dS/AdS metric, but first let us
examine the symmetries of the solutions.

\subsubsection{Symmetries of constant curvature solutions}

Constant curvature dS/AdS manifolds are maximally symmetric and have
three independent Killing vectors in 1+1 dimensions. In a flat 2+1
dimensional embedding space these generate the boosts and rotation
in the $SO(2,1)$ or $SO(1,2)$ symmetry group that preserves the dS
or AdS hyperboloid respectively. Two-dimensional dS and AdS are
related by interchange of the spacelike and timelike dimensions so
the corresponding solutions are closely related. We shall focus on
the dS case here, and indicate the corresponding results for AdS at
the end.

In terms of the Minkowski coordinates $X^0$, $X^1$, $X^2$ of the
flat 2+1 dimensional spacetime the generators of $SO(2,1)$ are the
two boosts $\textbf{K}_1 = X^0
\partial_1 + X^1
\partial_0$ and $\textbf{K}_2 = X^0 \partial_2 + X^2 \partial_0$, and
one rotation $\textbf{J} = X^1 \partial_2 - X^2 \partial_1$. These
form the Lie algebra with commutators
\bea [\textbf{K}_1, \textbf{K}_2] &=& \textbf{J} \nonumber\\
\left[\textbf{J}, \textbf{K}_1\right] &=& -\textbf{K}_2 \nonumber\\
\left[\textbf{J}, \textbf{K}_2\right] &=& \textbf{K}_1.
\label{so21alg}\eea
We are interested in the subgroup under which also the aether is
invariant. The corresponding Killing vectors are identical to the
Killing vectors of $\eta_{ab}$ that commute with $u'^a$, since ${\cal
L}_{\xi} \eta_{ab} = {\cal L}_{\xi} u'_a = 0$ implies ${\cal L}_{\xi}
g_{ab} = {\cal L}_{\xi} u_a = 0$. Their algebra is given by
(\ref{algebra1}). These Killing vectors must generate a two
dimensional non-Abelian subgroup of $SO(2,1)$. The only
two-dimensional non-Abelian subalgebras of (\ref{so21alg}) are
generated by a boost and a \textit{null rotation}, for example
\beq \left[\textbf{K}_1, \textbf{J}+\textbf{K}_2\right] =
\textbf{J}+\textbf{K}_2. \label{algebra2}\eeq
This coincides with the flat spacetime algebra (\ref{algebra1})
discussed in Section \ref{flatsol}, and so reveals the geometrical
nature of the symmetry group of our solutions.

Acting with the null rotation $\textbf{J}+\textbf{K}_2$ as a
differential operator one sees that the combination $X^0+X^1$ is
invariant. Thus, the flow lines of this null rotation on the
hyperboloid are the intersections of null planes $X^0+X^1 =
\rm{const.}$ with the embedded hyperboloid. We shall now reexpress
the dS solution in the ``planar" coordinate system adapted to the
generator of null rotations. This will help to illustrate the nature
of the aether field in this solution and exhibit which patch of dS
is covered by a nonsingular aether.

\subsubsection{$\b<1$: de Sitter solution in planar coordinates} \label{Planar}

In planar coordinates $(t,x)$ the unit dS hyperboloid
$(X^0)^2-(X^1)^2-(X^2)^2=-1$ is described by the embeddings
\bea X^0 &=& -\sinh t - \half x^2 e^{t} \nonumber\\
X^1 &=& -\cosh t + \half x^2 e^{t}  \nonumber \\
X^2 &=& x e^{t}\label{planar} \eea
\cite{Spradlin:2001pw}. Since $X^0 + X^1 = -e^{t}$, lines of
constant $t$ are the flow lines associated with the null rotation
discussed above. The full range of $t$ in $(-\infty,\infty)$
foliates half of the hyperboloid. Using $ds^2 =
(dX^0)^2-(dX^1)^2-(dX^2)^2$ with (\ref{planar}) the induced 1+1
dimensional metric on the hyperboloid is found to be
\beq ds^2 = dt^2 - e^{2t} dx^2. \eeq
In planar coordinates the null rotation symmetry generated by
$\partial_x$ is manifest.

The solution (\ref{constcurvmetric}) has curvature scalar given by
(\ref{R'}), whereas the unit hyperboloid has curvature scalar $-2$.
Hence the two agree when units are chosen so that the inverse length
$k$ is given by $k=(\b^{-1}-1)^{-1/2}$. Put differently, they agree
in units with the ``Hubble constant"
\beq H=k(\b^{-1}-1)^{1/2} \label{H}\eeq
equal to unity. We now adopt such units for notational brevity. The
results can be written in arbitrary units by inserting the
appropriate powers of $H$ to give each quantity the correct
dimension.

Under the coordinate transformation
\bea w&=&e^{t}\\
v&=&2\b^{-1/2} x-(\b^{-1}+1)e^{-t} \eea
the metric (\ref{constcurvmetric}) (in units with $H=1$) takes the
planar form
\beq ds^2 = dt^2 - e^{2t} dx^2, \eeq
and the aether (\ref{constcurvu}) takes the form
\beq
u=(1-\b)^{-1/2}\left(\partial_t-\b^{1/2}e^{-t}\partial_x\right).
\label{constcurvutx} \eeq
The flow lines of the aether are given by
\beq x-\sqrt{\b}e^{-t}={\rm const.}\label{planarflow} \eeq
The symmetries under which this aether is invariant are the null
rotation generated by the Killing vector $\partial_x$ and the boost
$w\partial_w-v\partial_v$ which in planar coordinates takes the form
$\partial_t -x\partial_x$. This is a combined time translation and
spatial contraction. In terms of the embedding coordinates (\ref{planar}),
the flow lines are given by the intersections of the planes
\beq \frac{X^2-\sqrt{\beta}}{X^0+X^1}={\rm const.}
\label{embedflow}\eeq
with the de Sitter hyperboloid.

\subsubsection{de Sitter solution in global coordinates}

To further visualize how the aether flow behaves and what part of de
Sitter spacetime it covers in a nonsingular manner, we transform to
global Robertson-Walker coordinates $(T,\phi)$, which in
two-dimensions arise from foliating the hyperboloid with circles.
These are related to the embedding coordinates $X^{0,1,2}$ of
(\ref{planar}) via
\bea X^0 &=& \sinh T \nonumber\\
X^1 &=&  \cosh T \cos \phi\nonumber\\
X^2 &=&\cosh T  \sin\phi, \label{RW}\eea
and they yield the line element
\beq ds^2 = dT^2 - \cosh^2 T\, d\phi^2. \eeq
In these coordinates only the rotation symmetry generated by
$\textbf{J}$ is manifest. The ranges $T\in(-\infty,\infty)$ and
$\phi\in(-\pi,\pi)$ cover the entire manifold.

\begin{figure}
 \includegraphics[width=10cm, height=5cm,angle=0]{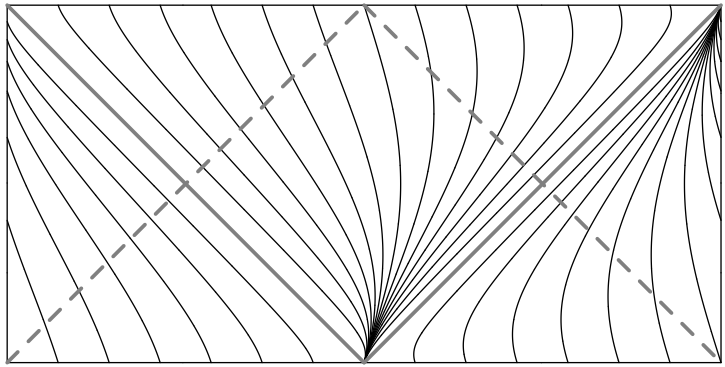}\\
\caption{\label{udS} Conformal diagram of 1+1 dS spacetime with the
flow lines of the aether field. Horizontal lines at the top and
bottom represent null infinity, while the two vertical lines at left
and right are identified. The solid and dashed grey lines form the
past and future horizons of the Killing horizon for the boost
symmetry under which the aether is invariant. The slope of the flow
lines is $-\b^{-1/2}$ on all boundaries of the diagram which is
drawn for the case $\b=0.1$.}
\end{figure}

If we introduce the new coordinate $\t$ via $\cosh T = \sec \t$, the
metric takes the conformally flat form
\beq ds^2 = \sec^2 \t[d\t^2 - d\phi^2],\eeq
and the finite range of $\t\in(-\pi/2,\pi/2)$ covers the entire
manifold. In these coordinates the flow lines (\ref{planarflow}) are
given by
\beq \frac{\sin\phi-\sqrt{\b}\cos\t}{\cos\phi+\sin\t}={\rm
const.}\label{globalflow} \eeq
The flow lines are plotted in Fig. \ref{udS}.
 The aether is regular in the planar coordinate system,
which covers the triangle with solid grey edges. On these edges
$u^a$ becomes infinitely stretched as it approaches a null
direction. The solid grey lines form the past horizon part of the
Killing horizon for the boost symmetry under which the aether is
invariant, while the dashed grey lines form the future horizon part.
The aether cannot possibly be regular at the bifurcation points
where the past and future horizons intersect, since these are fixed
points of the Killing flow hence a unit timelike vector cannot be
invariant there. (A similar circumstance occurs in the context of
the 3+1 dimensional black hole solutions in Einstein-aether
theory~\cite{Eling:2004dk,Eling:2006ec}.) However, the aether is
regular on the horizon to the future of the bifurcation points. This
solution therefore provides a setting with a nonsingular aether
flowing across a future horizon.

\subsubsection{$\beta>1$: Anti-de Sitter solution}

When $\b>1$ the curvature scalar (\ref{R'}) is positive, hence (with
our signature choice) the constant curvature solutions for this
theory correspond to anti-de Sitter space. In two dimensions dS and
AdS are exactly the same spacetime locally, only with a reversal in
the identification of what are the timelike and spacelike
directions. Rather than going through the details we simply remark
here that the aether solution for the AdS case can be obtained from
the dS case by interchanging the planar $t$ and $x$ coordinates.
This leads to the AdS metric in Poincar\'e coordinates, covering the
so-called ``Poincar\'e patch", and to the $u^a$ field appropriate to
the AdS space. The  flow lines of the aether are again given in the
embedding coordinates by (\ref{embedflow}), only now with $\b>1$. In
Fig. \ref{udS} we plot this flow and the Killing horizons in a
conformal diagram for AdS.
 To avoid closed timelike curves
we can pass to the covering space as is usually done, in which case
the diagram should be extended infinitely in the vertical direction.

\section{Discussion}

In this paper we have shown that the general Einstein-aether action
can be parameterized by two coupling constants in 1+1 dimensional
spacetime, and the classical equations of motion depend only on one
combination $\b$ of these. Hence there is a one-parameter family of
classical theories. Using a field redefinition of the metric, we
demonstrated that for $\b>0$ the theory can be reduced to a form
involving only one coupling constant which does not affect the
classical solutions. The only solutions to this reduced theory are a
flat metric together with one of three distinct types of solutions
for the aether field. Via the inverse field redefinition these
produce all solutions for the the generic theory, namely (i) flat
spacetime with constant aether, (ii) constant curvature spacetimes
with a uniformly accelerated $u^a$ invariant under a two-dimensional
symmetry group generated by a boost and a null rotation, and (iii) a
non-constant curvature spacetime that has no Killing symmetries and
contains singularities. The sign of the curvature is determined by
whether the coupling $\b$ is less or greater than one. For $\b<0$
only the solutions (i) and (iii) are present.

Unlike in dilaton gravity, there are no asymptotically flat black
hole solutions, although the de Sitter and anti-de Sitter solutions
possess Killing horizons that could allow issues of black hole
thermodynamics to be studied. This classical study of the behavior
of Einstein-aether theory in 1+1 dimensions may provide a starting
point for further investigations into semiclassical and fully
quantum toy models of quantum gravity with a dynamical preferred
frame.

\section*{Acknowledgements}

This work was supported in part by the NSF under grants PHY-0300710
and PHY-0601800 at the University of Maryland.

%\section{Appendix}\label{Appendix}
%\renewcommand{\theequation}{A.\arabic{equation}}
%\setcounter{equation}{0}

\end{document}